\documentclass{article}
\usepackage{spconf,amsmath,graphicx,hyperref}
\usepackage{amssymb}
\usepackage{cite}
\usepackage{bm} 

\usepackage{tikz}
\usepackage{tikzscale}
\usetikzlibrary{positioning, shapes.geometric, angles,arrows,intersections, calc, automata, shapes.misc, arrows.meta, decorations.pathreplacing, decorations.markings, quotes, bending,}

\usepackage{subcaption}
\usepackage{graphicx}
\usepackage[rightcaption]{sidecap}
\usepackage{gensymb}
\usepackage{multicol}
\usepackage[nolist]{acronym}
\usepackage{pgfplots}
\usepackage{pgf}
\usepgfplotslibrary{groupplots}
\pgfplotsset{compat=newest}
\usepackage{flushend}
\usepackage{balance}

\tikzset{cross/.style={cross out, draw=black, minimum size=2*(#1-\pgflinewidth), inner sep=0pt, outer sep=0pt}, 
cross/.default={3pt}}

\usepackage{booktabs,siunitx}

\usepackage[nolist]{acronym}

\begin{acronym}
\acro{stft}[STFT]{short-time Fourier transform}
\acro{doa}[DoA]{direction of arrival}
\acro{istft}[iSTFT]{inverse short-time Fourier transform}
\acro{dnn}[DNN]{deep neural network}
\acro{pesq}[PESQ]{perceptual evaluation of speech quality}
\acro{polqa}[POLQA]{perceptual objectve listening quality analysis}
\acro{wpe}[WPE]{weighted prediction error}
\acro{psd}[PSD]{power spectral density}
\acro{rir}[RIR]{room impulse response}
\acro{snr}[SNR]{signal-to-noise ratio}
\acro{lstm}[LSTM]{long short-term memory}
\acro{polqa}[POLQA]{Perceptual Objectve Listening Quality Analysis}
\acro{sdr}[SDR]{signal-to-distortion ratio}
\acro{estoi}[ESTOI]{extended short-term objective intelligibility}
\acro{elr}[ELR]{early-to-late reverberation ratio}
\acro{tcn}[TCN]{temporal convolutional network}
\acro{rls}[RLS]{recursive least squares}
\acro{asr}[ASR]{automatic speech recognition}
\acro{ha}[HA]{hearing aid}
\acro{ci}[CI]{cochlear implant}
\acro{mac}[MAC]{multiply-and-accumulate}
\acro{vae}[VAE]{variational auto-encoder}
\acro{gan}[GAN]{generative adversarial network}
\acro{tf}[T-F]{time-frequency}
\acro{sde}[SDE]{stochastic differential equation}
\acro{ode}[ODE]{ordinary differential equation}
\acro{drr}[DRR]{direct to reverberant ratio}
\acro{lsd}[LSD]{log spectral distance}
\acro{sisdr}[SI-SDR]{scale-invariant signal-to-distortion ratio}
\acro{mos}[MOS]{mean opinion score}
\acro{map}[MAP]{maximum a posteriori}
\acro{rtf}[RTF]{real-time factor}
\acro{ema}[EMA]{exponential moving average}
\acro{rkhs}[RKHS]{reproducible kernel Hilbert space}
\acro{mmd}[MMD]{maximum mean discrepancy}
\acro{ica}[ICA]{independence component analysis}
\acro{mvdr}[MVDR]{minimum variance distortionless response}
\acro{pf}[PF]{postfilter}
\acro{tdoa}[TdoA]{time difference of arrival}
\acro{atf}[ATF]{acoustic transfer function}
\acro{pf}[PF]{post-filter}
\acro{di}[DI]{directivity index}
\acro{lr}[LR]{learning rate}
\acro{ssl}[SSL]{sound source localization}
\acro{ipd}[IPD]{inter-channel phase difference}
\acro{irm}[IRM]{ideal ratio mask}
\acro{scm}[SCM]{spatial covariance matrix}
\end{acronym}

\title{An Analysis of Joint Nonlinear Spatial Filtering \\for Spatial Aliasing Reduction}

\name{Alina Mannanova, Jakob Kienegger, Timo Gerkmann \thanks{This work was supported by the Deutsche Forschungsgemeinschaft (DFG, German Research Foundation) under grant 508337379.}}
\address{Signal Processing (SP), University of Hamburg, Hamburg, Germany}

\begin{document}
\ninept
\maketitle
\begin{abstract}
The performance of traditional linear spatial filters for speech enhancement is constrained by the physical size and number of channels of microphone arrays. For instance, for large microphone distances and high frequencies, spatial aliasing may occur, leading to unwanted enhancement of signals from non-target directions. Recently, it has been proposed to replace linear beamformers by nonlinear deep neural networks for joint spatial-spectral processing. While it has been shown that such approaches result in higher performance in terms of instrumental quality metrics, in this work we highlight their ability to efficiently handle spatial aliasing. In particular, we show that joint spatial and tempo-spectral processing is more robust to spatial aliasing than traditional approaches that perform spatial processing alone or separately with tempo-spectral filtering. The results provide another strong motivation for using deep nonlinear networks in multichannel speech enhancement, beyond their known benefits in managing non-Gaussian noise and multiple speakers, especially when microphone arrays with rather large microphone distances are used. 

\end{abstract}
\section{Introduction}
\label{sec:intro}

Microphone arrays are widespread in consumer electronics, from smartphones and tablets to smart speakers and hearing aids. The key advantage of using multiple microphones over a single-microphone setup is the ability to leverage spatial filtering, which allows devices to focus on sound coming from a specific \ac{doa} while suppressing interference and noise from other directions \cite{vary23}. The effectiveness of such spatial filtering depends on the distance between microphones in the array. 

Just as time-domain signals are sampled at discrete intervals, microphones in the array discretely sample the sound field across space, with spatial information encoded in the \ac{tdoa} \cite{vary23}. In analogy to the Nyquist sampling theorem for time-domain signals that depends on the sampling period, spatial sampling follows a Nyquist criterion that depends on the inter-microphone distance \cite{johnson1992array_concepts}. When such Nyquist criterion is violated, aliasing occurs, creating ambiguity. While temporal aliasing results in frequency ambiguity, \textit{spatial aliasing} leads to \ac{doa} ambiguity  meaning that array has problems to distinguish between signals arriving from different directions \cite{vary23}. This spatial aliasing can be avoided by placing the microphones at the same distance or closer than half of the shortest wavelength present in the signal, as stated by the spatial Nyquist criterion \cite{Trees2002OptimumArray}. Since this criterion depends on wavelength, spatial aliasing is frequency-dependent: higher frequencies with shorter wavelengths are more prone to aliasing for a given inter-microphone distance \cite{capon1969wavenumbers, dmochowski2009on_aliasing}.

Different applications demand different array designs. In compact devices like hearing aids where closely spaced microphones are often placed on the same ear (typically \SI{1}{\centi\meter} apart \cite{Hendriks2011EstNoiseCorMx}), spatial aliasing would occur only from \SI{17}{\kilo\hertz}, enabling reliable spatial filtering across most frequencies but with limited spatial resolution. In contrast, larger arrays like those used in binaural hearing aids with microphones on both ears (\SI{17}{\centi\meter} distance \cite{doclo2016hearing_aids}) can cover wider area and offer improved spatial resolution \cite{lombard_kellermann2011tdoa} but become susceptible to aliasing above \SI{1}{\kilo\hertz}. This illustrates a fundamental trade-off in array design: smaller inter-microphone distance prevents high-frequency aliasing but limits directivity, while larger distance improves spatial resolution but introduces aliasing risks \cite{gosswein2017stereophonic}.

Approaches to address this trade-off include irregularly spaced arrays \cite{johnson1992array_concepts,yang2021freq_diff} or increasing the number of microphones \cite{tang2011ssr}, but these are not always practical due to size or complexity constraints, making alternative methods to mitigate spatial aliasing necessary.

Several studies \cite{yang2021freq_diff, tang2011ssr,  reddy2014UnambiguousSpeechDOA, reddy2015doa, chen2017ipd2steps, song21ipd_allfreq} have addressed the \ac{doa} ambiguities caused by spatial aliasing when ideal inter-microphone distance according to Nyquist criterion is not possible. These approaches exploit the frequency-dependent nature of spatial aliasing through different strategies. \cite{tang2011ssr} compares aliasing patterns across two sufficiently spaced frequencies such that only the true \ac{doa} remains consistent, a principle also used by \cite{yang2021freq_diff} with a frequency-difference algorithm. Similarly, \cite{reddy2014UnambiguousSpeechDOA} and \cite{reddy2015doa} sequentially estimate \ac{doa} beginning with low frequencies less affected by aliasing to guide estimation in higher frequencies that provide better angular resolution. Alternatively, \cite{chen2017ipd2steps} and \cite{song21ipd_allfreq} leverage multiple \ac{ipd} hypotheses and select the consistent one across frequencies to resolve aliasing ambiguities. More recently, \cite{guzik2025unet_aliasing} proposed a DNN-based de-aliasing filter for conventional beamforming.
 
While existing approaches to spatial aliasing focus on resolving \ac{doa} ambiguities for localization tasks or employ hybrid DNN-beamforming strategies, this work takes a different path by exploring whether nonlinear joint spatial tempo-spectral processing can directly perform effective speech enhancement on arrays with inter-microphone distances exceeding the Nyquist criterion, without explicitly resolving spatial ambiguities.

\section{Background}
\label{sec:background}

\subsection{Signal model}
\label{ssec:signal_model}
To enable spatial filtering for speaker extraction, we consider a microphone array with $M$ channels recording a target speech signal $s(t)$ in a noisy reverberant environment. Each microphone signal $y_m(t)$ at time index $t$ is a mixture of a delayed reverberant version of the target signal $x_m(t)$ and additive noise $n_m(t)$. The corresponding frequency-domain representation using the \ac{stft} at frequency bin $k$ and time frame $l$ is given by
\begin{equation}
\label{eq:signal_model_stft}
    Y_m(k, l) = X_m(k, l) + N_m(k, l), \quad m = 1,\ldots,M
\end{equation}
where $X_m(k, l)$ and $N_m(k, l)$ are the \ac{stft} coefficients of the recorded target signal and noise, respectively. The goal of speaker extraction is to estimate the clean target signal $\widehat{S}(k, l)$ coming from \ac{doa} $\theta_s$ using these noisy multichannel recordings.

\subsection{Traditional linear spatial beamforming}
\label{ssec:beamforming}
A traditional approach to achieve this speaker extraction is linear beamforming. This technique combines microphone signals to create a directional beam that enhances sound from target location while suppressing interference from others \cite{vary23}.

Among various beamforming techniques, the \ac{mvdr} beamformer is commonly used \cite{benesty2023arrays}. It minimizes the output power while maintaining an undistorted response toward the target look direction. The optimal \ac{mvdr} filter coefficients are frequency-dependent. They are computed using the noise correlation matrix $\bm{\Phi}_{k, l}^{n} \in \mathbb{C}^{M \times M}$ and the steering vector $\bm{a}_{k}(\Delta\tau_m) \in \mathbb{C}^M$, where $\Delta\tau_m$ denotes the time delays relative to the reference microphone that depend on the target source \ac{doa} $\theta_s$, as given in \cite{vary23}
\begin{equation}
    \label{eq:mvdr}
    \bm{h}_{\textrm{MVDR}} = \frac{(\bm{\Phi}^{n})^{-1}\bm{a}}{\bm{a}^H(\bm{\Phi}^{n})^{-1}\bm{a}},
\end{equation}
where $\cdot^H$ denotes the Hermitian transpose and time-frequency indices are omitted for clarity. The target signal estimate is obtained as $\widehat{\bm{S}} = \bm{h}^H_{\textrm{MVDR}}\bm{Y}$, where the \ac{mvdr} beamformer applies a linear transformation to the noisy microphone signals.

Despite their popularity, linear spatial beamformers face certain limitations. Previous research \cite{tesch2021nonlinear} has proven that the \ac{mvdr} beamformer is not a sufficient statistics for estimating clean speech in non-Gaussian noise and can suppress no more than $M-1$ sources \cite{vary23}. Furthermore, as demonstrated in \cite{vary23, Trees2002OptimumArray}, the design of microphone arrays for linear spatial filtering requires careful attention to the distance between microphones. This study focuses specifically on this latter aspect.

\subsection{Spatial aliasing and its impact on beamforming}
\label{ssec:aliasing}

\begin{figure}[t]
  \centering
  \resizebox{0.2\textwidth}{!}{\begin{tikzpicture}

\def\angleValue{60}
\def\micRadius{4pt}  %

\draw (0,0) -- (4.75,0);

\coordinate (mic1base) at (0.3,0);
\coordinate (mic2base) at (3,0);
\coordinate (mic1) at (0.3,-\micRadius);
\coordinate (mic2) at (3,-\micRadius);
\draw[thick] (mic1) circle[radius=\micRadius];
\draw[thick] (mic2) circle[radius=\micRadius];
\draw[thick] ($(mic1base)+(-0.15,0)$) -- ($(mic1base)+(0.15,0)$);
\draw[thick] ($(mic2base)+(-0.15,0)$) -- ($(mic2base)+(0.15,0)$);

\draw[dotted, thick] (mic1base) -- +(\angleValue:2.3);
\draw[dotted, thick] (mic2base) -- +(\angleValue:2.3);
\draw (mic2base) -- +(\angleValue+90:3);

\coordinate (mic1right) at ($(mic1base)+(1,0)$);
\coordinate (mic1up) at ($(mic1base)+(\angleValue:1)$);
\pic [draw, angle radius=4mm, "$\theta_s$" {shift={(0.35,0.25)}}] {angle = mic1right--mic1base--mic1up};

\coordinate (mic2right) at ($(mic2base)+(1,0)$);
\coordinate (mic2up) at ($(mic2base)+(\angleValue:1)$);
\pic [draw, angle radius=4mm, "$\theta_s$" {shift={(0.35,0.25)}}] {angle = mic2right--mic2base--mic2up};

\coordinate (startdistance) at ($(mic1)-(0,0.4)$);
\coordinate (enddistance) at ($(mic2)-(0,0.4)$);
\draw[stealth-stealth] (startdistance) -- node[below] {$d$} (enddistance);
\draw (startdistance) ++ (0,-6pt) -- ++ (0,12pt);
\draw (enddistance) ++ (0,-6pt) -- ++ (0,12pt);

\node at (4.2,2.6) {$\bm{s}$};

\node[below right=1pt and 3pt of mic2] {$\bm{y}_1$};
\node[below left=1pt and 3pt of mic1] {$\bm{y}_2$};

\draw[decorate,decoration={brace,amplitude=5pt}] 
    ($(mic1base)$) -- ($(mic1base)+(\angleValue:1.35)$) node[midway,above,sloped,yshift=5pt] {$\Delta r_2$};

\end{tikzpicture}}
  \caption{Visualization of wave propagation (dotted line) from a source $\bm{s}$ located at \ac{doa} $\theta_s$ in a free-field microphone array with inter-microphone distance $d$.}
  \label{fig:waves}
\end{figure}
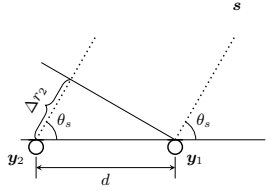

For spatial filtering, the key information lies in the relative time delays between microphone signals caused by the target source. To illustrate this concept, we consider the setup depicted in Fig. \ref{fig:waves} which shows two microphones separated by a distance $d$. A source $\bm{s}$ located in the far field generates a plane wave of frequency $f_k$ corresponding to the $k$-th frequency bin. This wave, shown as a dotted line, arrives at an angle $\theta_s$ relative to the array axis. As a result, there is a path difference $\Delta r_2 = d\cos{\theta_s}$ between the wave reaching two microphones which causes \ac{tdoa} $\Delta\tau_2 = d\cos\theta_s/c$, where $c$ is the speed of sound. Neglecting the source-array propagation time, the time domain microphone signals can be expressed as $y_1(t) = s(t)$ and $y_2(t) = y_1(t - \Delta \tau_2)$. 

In the frequency domain, this time delay results in a linear phase shift $\Delta \phi_m(k)$ \cite{oppenheim99dsp} and the signal of the second microphone can be formulated as $Y_2(k, l) = Y_1(k, l)e^{-j\Delta\phi_2(k)}$ with
\begin{equation}
\label{eq:phase_shift}
    \Delta \phi_2(k) = 2\pi f_k \Delta \tau_2 = \frac{2\pi f_k d\cos{\theta_s}}{c}.
\end{equation}

Eq. \ref{eq:phase_shift} shows that the phase shift varies with frequency and when applied as $e^{j\Delta\phi_m(k)}$ is ambiguous in multiples of $2\pi$, resulting in spatial aliasing. This means that multiple distinct signal directions can produce identical phase differences between the array elements, leading to ambiguity in \ac{doa} \cite{vary23}.

To avoid this spatial aliasing, it is required that the maximum possible phase shift produced by highest frequency $f_{\mathrm{max}}$ present in the signal is less than $\pi$ \cite{vary23}
\begin{equation}
\label{eq:aliasing_condition}
\frac{2\pi f_{\mathrm{max}} d}{c} \leq \pi \Rightarrow d \leq \frac{c}{2f_\mathrm{max}}.
\end{equation}

This gives us the spatial aliasing cut-off frequency $f_\mathrm{a}$ \cite{chen2017ipd2steps}, which is the maximum frequency that can be processed without spatial aliasing for given $d$
\begin{equation}
\label{eq:spatial_aliasing_freq}
f_\mathrm{a} = \frac{c}{2d}.
\end{equation}

Therefore, for signals with maximum frequency $f_\mathrm{max}$, the inter-microphone distance must satisfy $d \leq \lambda_{\mathrm{min}}/2$ to ensure one-to-one mapping between the \ac{doa} and phase difference, where $\lambda_{\mathrm{min}} = c / f_{\mathrm{max}}$ is the shortest wavelength corresponding to the highest frequency component.

\section{DNN-based multichannel signal processing}
\label{sec:dnn}

Traditional beamforming techniques employ linear spatial filtering applied to each frequency bin separately without cross-frequency interactions. These approaches operate either independently or in combination with a spectral (non)linear \ac{pf}. In contrast, \ac{dnn}-based end-to-end multichannel filters like FT-JNF \cite{Tesch2023Insights} or SpatialNet \cite{SpatialNet} enable simultaneous nonlinear processing in both tempo-spectral and spatial domains. This joint processing approach typically involves two key stages. First, cross-band processing focuses on spatial and spectral information by processing multichannel data across frequency bins to capture correlations between frequencies. The following narrow-band processing block focuses on spatial and temporal information, analyzing patterns within individual frequency bins independently.

Recent studies \cite{Tesch2023Insights, briegleb2024analysis, cohen2025explainable} have analyzed the behavior and characteristics of \ac{dnn}-based multichannel filters to better understand their processing mechanisms. Similarly, in this study we examine the speech enhancement and spatial selectivity characteristics of end-to-end joint nonlinear filters with respect to different inter-microphone distances, including a configuration that violates the spatial Nyquist criterion.

\section{Experimental analysis}
\label{sec:exp_setup}

\subsection{Microphone array setup and processing pipelines}
\label{ssec:array_pipelines}

We consider an array of two microphones that are positioned at $d = \SI{1}{\centi\meter}$ and $d = \SI{17}{\centi\meter}$ from each other. These configurations serve as illustrative examples of inter-microphone distances for a two-microphone hearing aid when the microphones are placed either on the same ear \cite{Hendriks2011EstNoiseCorMx} or on both ears \cite{doclo2016hearing_aids}, assuming a simplified model without head effects. The different distances $d$ lead to different spatial aliasing frequencies, which determine the frequency range over which linear beamformers can operate without aliasing. Considering the speech signals up to \SI{8}{\kilo\hertz}, the smaller distance ($d = \SI{1}{\centi\meter}$) yields a spatial aliasing frequency of about \SI{17}{\kilo\hertz} according to Eq. \ref{eq:spatial_aliasing_freq}, which lies well above our frequency range of interest. In contrast, the larger distance ($d = \SI{17}{\centi\meter}$) reduces the aliasing frequency to \SI{1}{\kilo\hertz}, placing it inside the considered spectrum.

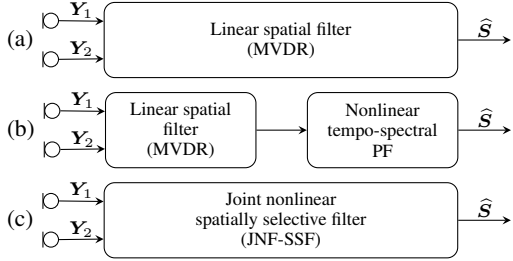
\begin{figure}
\centering
  \begin{subfigure}{0.4\textwidth}
    {\phantomsubcaption{}\label{fig:mvdr}}
     \tikz\node[inner sep=0pt,label={west:(\subref{fig:mvdr})}] 
     {\begin{tikzpicture}[
font=\footnotesize,
sy+/.style = {yshift= 2.5mm}, 
sy-/.style = {yshift=-2.5mm},
]
\tikzstyle{every node}=[font=\scriptsize]
\centering
    \node[circle, draw=black, 
			minimum width=0.2cm, 
		minimum height=0.2cm] at (-9.5, 1.2)(mic1){};
    \draw ([yshift=3pt]mic1.west) -- ([yshift=-3pt]mic1.west);

    \node[circle, draw=black, below=0.3cm of mic1,
		minimum width=0.2cm, 
		minimum height=0.2cm] (mic2) {};
    \draw ([yshift=3pt]mic2.west) -- ([yshift=-3pt]mic2.west);

    \coordinate(mid_mic) at ($(mic1)!0.5!(mic2)$);

    \node[rounded corners, draw=black, align=center, 
		minimum width=4.7cm, 
		minimum height=1cm,
        right= 0.7cm of mid_mic,
        ] (mvdr) {Linear spatial filter \\ (MVDR)};

    \draw[draw=black, -stealth] (mic1.east) -- ([sy+] mvdr.west) node[midway, above, yshift=1pt] {$\bm{Y}_1$};
    \draw[draw=black, -stealth]  (mic2.east) -- ([sy-] mvdr.west) node[midway, below, yshift=7pt] {$\bm{Y}_2$};

    \draw[draw=black, -stealth]  (mvdr.east) -- ++ (0.7,0) node[midway, above, yshift=1pt] {$\bm{\widehat{S}}$}; 
\end{tikzpicture}};
     \vspace*{1.5mm}
  \end{subfigure} 
  
  \begin{subfigure}{0.4\textwidth}
    {\phantomsubcaption{}\label{fig:mvdr_pf}}
     \tikz\node[inner sep=0pt,label={west:(\subref{fig:mvdr_pf})}] 
     {\begin{tikzpicture}[
font=\footnotesize,
sy+/.style = {yshift= 2.5mm}, 
sy-/.style = {yshift=-2.5mm},
]
\tikzstyle{every node}=[font=\scriptsize]
\centering
    \node[circle, draw=black, 
			minimum width=0.2cm, 
		minimum height=0.2cm] at (-9.5, 1.2)(mic1){};
    \draw ([yshift=3pt]mic1.west) -- ([yshift=-3pt]mic1.west);

    \node[circle, draw=black, below=0.3cm of mic1,
		minimum width=0.2cm, 
		minimum height=0.2cm] (mic2) {};
    \draw ([yshift=3pt]mic2.west) -- ([yshift=-3pt]mic2.west);

    \coordinate(mid_mic) at ($(mic1)!0.5!(mic2)$);

    \node[rounded corners, draw=black, align=center, 
		minimum width=2cm, 
		minimum height=1cm,
        right= 0.7cm of mid_mic,
        ] (FT-JNF) {Linear spatial \\ filter \\ (MVDR)};
    \node[rounded corners, draw=black, align=center, 
		minimum width=2cm, 
		minimum height=1cm,
        right= 0.67cm of FT-JNF,
        ] (PF) {Nonlinear \\ tempo-spectral \\ PF};

    \draw[draw=black, -stealth] (mic1.east) -- ([sy+] FT-JNF.west) node[midway, above, yshift=1pt] {$\bm{Y}_1$};
    \draw[draw=black, -stealth]  (mic2.east) -- ([sy-] FT-JNF.west) node[midway, below, yshift=7pt] {$\bm{Y}_2$};
    \draw[draw=black, -stealth]  (FT-JNF.east) -- (PF.west) node[midway, below] {};  %

    \draw[draw=black, -stealth]  (PF.east) -- ++ (0.7,0) node[midway, above, yshift=1pt] {$\bm{\widehat{S}}$}; %
\end{tikzpicture}};
     \vspace*{1.5mm}
  \end{subfigure}
  
  \begin{subfigure}{0.4\textwidth}
    {\phantomsubcaption{}\label{fig:ftjnf}}
     \tikz\node[inner sep=0pt,label={west:(\subref{fig:ftjnf})}] 
     {\begin{tikzpicture}[
font=\footnotesize,
sy+/.style = {yshift= 2.5mm}, 
sy-/.style = {yshift=-2.5mm},
]
\tikzstyle{every node}=[font=\scriptsize]
\centering
    \node[circle, draw=black, 
			minimum width=0.2cm, 
		minimum height=0.2cm] at (-9.5, 1.2)(mic1){};
    \draw ([yshift=3pt]mic1.west) -- ([yshift=-3pt]mic1.west);

    \node[circle, draw=black, below=0.3cm of mic1,
		minimum width=0.2cm, 
		minimum height=0.2cm] (mic2) {};
    \draw ([yshift=3pt]mic2.west) -- ([yshift=-3pt]mic2.west);

    \coordinate(mid_mic) at ($(mic1)!0.5!(mic2)$);

    \node[rounded corners, draw=black, align=center, 
		minimum width=4.7cm, 
		minimum height=1cm,
        right= 0.7cm of mid_mic,
        ] (FT-JNF) {Joint nonlinear \\ spatially selective filter \\ (JNF-SSF)};

    \draw[draw=black, -stealth] (mic1.east) -- ([sy+] FT-JNF.west) node[midway, above, yshift=1pt] {$\bm{Y}_1$};
    \draw[draw=black, -stealth]  (mic2.east) -- ([sy-] FT-JNF.west) node[midway, below, yshift=7pt] {$\bm{Y}_2$};

    \draw[draw=black, -stealth]  (FT-JNF.east) -- ++ (0.7,0) node[midway, above, yshift=1pt] {$\bm{\widehat{S}}$}; 
\end{tikzpicture}};
  \end{subfigure}
  
  \caption{Multichannel signal processing pipelines, where $\bm{Y}_1$ and $\bm{Y}_2$ denote the multichannel recorded signals and $\bm{\widehat{S}}$ represents the enhanced target signal. (\subref{fig:mvdr}) Linear spatial filter (\subref{fig:mvdr_pf}) Linear spatial filter followed by a tempo-spectral \ac{pf} (\subref{fig:ftjnf}) Joint nonlinear spatially-selective tempo-spectral filter.}
  \label{fig:setups}
\end{figure}

To evaluate the influence of joint spatial and tempo-spectral nonlinear processing on the spatial aliasing problem, we compare this approach to methods based on a linear spatial filter. We examine speech enhancement and spatial selectivity performance for three pipelines illustrated in Fig. \ref{fig:setups}: an \ac{mvdr} beamformer used alone (\ref{fig:mvdr}), the \ac{mvdr} concatenated with a nonlinear \ac{pf} (\ref{fig:mvdr_pf}) and an end-to-end \ac{dnn}-based architecture JNF-SSF (\ref{fig:ftjnf}) which performs joint spatial and tempo-spectral filtering. The JNF-SSF architecture \cite{tesch2024ssf} extends FT-JNF \cite{Tesch2023Insights} by incorporating a target source direction as an additional input, enabling steering of the filter. To allow for a fair comparison in terms of the neural network capacity, FT-JNF, which shares the same base architecture as JNF-SSF, is adapted here to operate as a single-channel tempo-spectral \ac{pf} for the \ac{mvdr} beamformer by setting the number of microphones to one. A detailed description of FT-JNF and JNF-SSF architectures is provided in \cite{Tesch2023Insights, tesch2024ssf}. For all pipelines, the input signals are the two microphone signals $\bm{Y}_1$ and $\bm{Y}_2$, with the output given by the estimated target speech signal $\bm{\widehat{S}}$.

\subsection{Speech dataset and training parameters}
\label{ssec:setup_speech}

We use \texttt{pyroomacoustics} \cite{pyroomacoustics} to simulate training and evaluation datasets. Our experimental setup employs two speakers as distinct directional sources, meeting the source-to-microphone ratio criterion to ensure that \ac{mvdr} beamformer and JNF-SSF are evaluated under comparable conditions \cite{tesch2021nonlinear}. Similar to \cite{Fallahi2021SittedSpeaker}, the target and interfering speakers are seated with the microphone array placed at the same height of \SI{1.3}{\meter}, corresponding to ear level. Both speakers are positioned between \num{0} and $\pi$ with respect to the array to avoid front-back ambiguity with a minimum angular separation of $\pi/6$. For each utterance, we randomly sample room characteristics from Table \ref{tab:room} and microphone array placement within it, then choose the source-to-array distance between \qtyrange{1}{2}{\meter}. The clean speech signals are sourced from the WSJ0 corpus \cite{WSJ0}, with no overlap between training, validation, and testing samples. To ensure a valid comparison across examined array configurations, all dataset parameters are maintained, except for the inter-microphone distance. Thus, two distinct datasets are simulated with inter-microphone distance $d = \SI{1}{\centi\meter}$ and $d = \SI{17}{\centi\meter}$.

The loss function is taken from \cite{Tesch2023Insights}, while the number of training and evaluation samples per each target \ac{doa} follows \cite{tesch2023doa_ftjnf}. Target speaker positions $\theta_s$ are uniformly distributed across \num{60} angles from \num{0} to $\pi$. For all setups that include training a \ac{dnn} (\ac{pf} at the output of MVDR beamformer and JNF-SSF as a standalone multichannel filter), we use the Adam optimizer with an initial \ac{lr} of \num{e-3}. The \ac{lr} is dynamically adjusted, decreasing by a factor of \num{0.8} if the validation loss shows no improvement over three consecutive epochs. Training for all models is governed by an early stopping, terminating the process after ten epochs of stagnant validation performance. 

\begin{table}[th]
  \caption{Ranges of room characteristics for speech dataset}
  \label{tab:room}
  \centering
  \begin{tabular}{ c c c c }
    \toprule
    \multicolumn{1}{c}{\textbf{Width [m]}} & 
    \multicolumn{1}{c}{\textbf{Length [m]}} &
    \multicolumn{1}{c}{\textbf{Height [m]}} &
    \multicolumn{1}{c}{\textbf{$\mathbf{T}_{60}$ [s]}} \\
    \midrule
    $3 - 6$    &$3 - 9$    &$2.2 - 3.5$ &$0.2 - 0.5$~~~ \\
    \bottomrule
  \end{tabular}
\end{table}

\subsection{Spatial selectivity evaluation setup}
\label{ssec:setup_beampatterns}

To evaluate spatial selectivity of the three methods, we follow \cite{Tesch2023Insights} using spectrally white noise emitted from $-\pi$ to $\pi$ as target signals. This broadband signal enables assessment of filter responses across the full spectrum and allows to have frequency interactions important for JNF-SSF architecture. Additionally, testing \acp{dnn} on signals spectrally unseen during training while preserving spatial characteristics allows evaluation of spatial processing independently of tempo-spectral features. The \ac{mvdr} beamformer is optimized for a diffuse noise field. For each direction, multichannel filters produce estimated \acp{stft} from which we calculate time-averaged spectral amplitudes in \unit{\deci\bel}.

Across all angles, room dimensions are kept the same and set to median values from Table \ref{tab:room}, with the array at room center and source-array distance of \SI{1.5}{\meter}. Microphone array and source heights remain as described in \ref{ssec:setup_speech}. We simulate two datasets in an anechoic chamber to eliminate reflections and isolate directional responses, varying the inter-microphone distance $d$ between datasets as in \ref{ssec:setup_speech}.

\section{Results and discussion}
\label{sec:results}

\subsection{Impact of inter-microphone distance on speech enhancement}
\label{ssec:results_speech}

\begin{table*}[t]
\centering
\caption{Speaker extraction results for different inter-microphone distance $d$ with varying target speaker positions. Results show mean improvement ($\Delta$PESQ, $\Delta$SI-SDR) and absolute values (ESTOI) with 95\% confidence intervals. The best results are indicated in bold.}
\label{tab:mvdr_ssf_results_speech}
\sisetup{
    reset-text-series = false, 
    text-series-to-math = true, 
    mode=text,
    tight-spacing=true,
    round-mode=places,
    round-precision=2,
    table-format=+2.2,
    table-number-alignment=center,
    table-text-alignment=center,
    table-column-width = 6em
}
\begin{tabular}{l@{\hspace{2em}}
S[round-precision=2,table-format=2.2, table-alignment-mode=none]
@{\hspace{0.1em}}
S[round-precision=2,table-format=1.3, table-alignment-mode=none]
@{\hspace{3em}}
S[round-precision=2,table-format=2.1, table-alignment-mode=none]
@{\hspace{0.1em}}
S[round-precision=2,table-format=2.1, table-alignment-mode=none]
@{\hspace{3em}}
S[round-precision=1,table-format=2.1, table-alignment-mode=none]
@{\hspace{0.1em}}
S[round-precision=1,table-format=2.1, table-alignment-mode=none]}
    \toprule
    & \multicolumn{2}{c}{$\Delta$ PESQ $\uparrow$}
    & \multicolumn{2}{c}{ESTOI $\uparrow$}
    & \multicolumn{2}{c}{$\Delta$ SI-SDR [dB] $\uparrow$}\\
    \cmidrule(r{2.9em}){2-3}\cmidrule(lr{2.9em}){4-5}\cmidrule(l{0.9em}){6-7}
    & {$d = 1\mathrm{cm}$}& {$d = 17\mathrm{cm}$}
    &{$d = 1\mathrm{cm}$} & {$d = 17\mathrm{cm}$}
    & {$d = 1\mathrm{cm}$}& {$d = 17\mathrm{cm}$}\\
    \midrule
    MVDR       & 0.06$\pm$0.01  &  0.03$\pm$0.01
               & 0.49$\pm$0.01  &  0.47$\pm$0.01
               & 3.6$\pm$0.5   &  3.2$\pm$0.5  \\

    MVDR+PF         & 0.19$\pm$0.02   &  0.29$\pm$0.02
                    & 0.55$\pm$0.01  &  0.59$\pm$0.01
                    & 5.9$\pm$0.5  &  7.3$\pm$0.6  \\

    JNF-SSF  & \bfseries 0.78$\pm$0.03 & \bfseries 0.75$\pm$0.03 
            & \bfseries 0.73$\pm$0.01 & \bfseries 0.73$\pm$0.01
            & \bfseries 11.1$\pm$0.4 &  \bfseries 10.9$\pm$0.5 \\
    \bottomrule
\end{tabular}
\end{table*}

We examine the impact of inter-microphone distance $d$ on speech enhancement performance across different multichannel pipelines. Table \ref{tab:mvdr_ssf_results_speech} presents the mean improvement in \ac{pesq} and \ac{sisdr}, absolute values for \ac{estoi} for three approaches: \ac{mvdr} beamformer, \ac{mvdr} with \ac{pf} (MVDR + PF) and the \ac{dnn}-based end-to-end filter JNF-SSF. Although the \ac{mvdr} beamformer has access to oracle \ac{doa} and \ac{irm} for noise \ac{scm} estimation, it shows minimal improvement in \ac{pesq} and \ac{estoi} but achieves approximately \SI{3}{\deci\bel} improvement in \ac{sisdr} for both distances with $d=\SI{1}{\centi\meter}$ performing better. The addition of a \ac{pf} significantly enhances performance, particularly for the larger inter-microphone distance. JNF-SSF substantially outperforms both linear spatial filter-based frameworks, achieving nearly \SI{11}{\deci\bel} improvement in \ac{sisdr} while maintaining consistent performance across inter-microphone distances.

\subsection{Impact of inter-microphone distance on spatial selectivity}
\label{ssec:results_spatial}

\begin{figure}[h]
\includegraphics[width=\columnwidth]{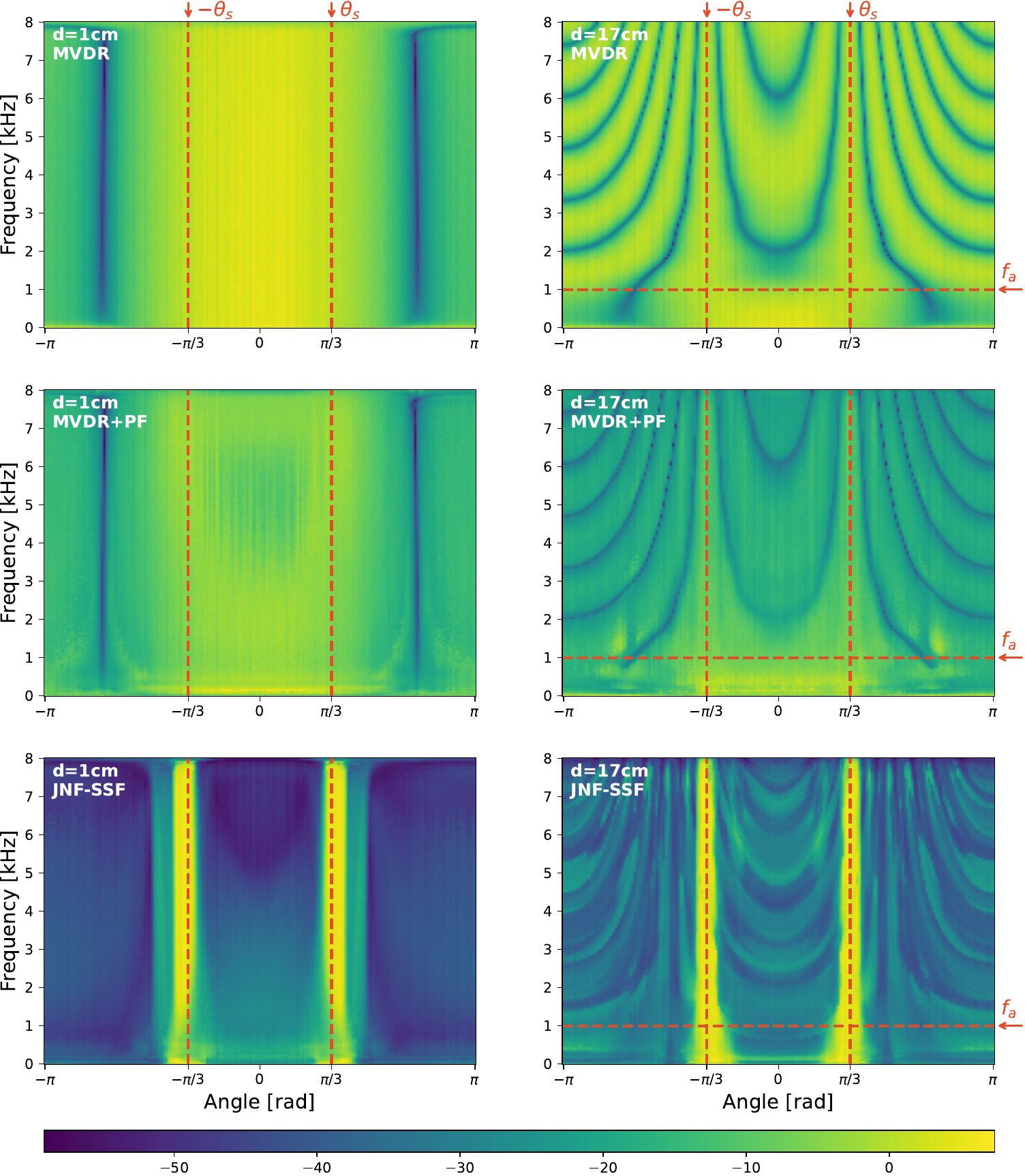}
\caption{Spatial selectivity comparison for $d = \SI{1}{\centi\meter}$ (left) and $d = \SI{17}{\centi\meter}$ (right): \ac{mvdr}(top), \ac{mvdr}+PF (middle) and JNF-SSF (bottom). The x-axis shows the angle of arrival, the y-axis represents frequency. Vertical dashed lines indicate the look-direction
$\theta_s = \pm\pi/3$, while the horizontal dashed line shows the aliasing cut-off frequency $f_\mathrm{a}$.}
    \label{fig:beampatterns_mvdr_ssf}
\end{figure}

Fig. \ref{fig:beampatterns_mvdr_ssf} shows the spatial selectivity patterns for the same multichannel pipelines: \ac{mvdr} (top), \ac{mvdr}+\ac{pf} (middle), and JNF-SSF (bottom). The left column shows results for $d = \SI{1}{\centi\meter}$, while the right column presents $d = \SI{17}{\centi\meter}$. In each subplot, the x-axis represents the angle of arrival and the y-axis shows frequency. Vertical dashed lines indicate the look-direction $\theta_s = \pm\pi/3$, and for $d = \SI{17}{\centi\meter}$, the horizontal dashed line marks the calculated aliasing cut-off frequency $f_\mathrm{a}$ according to Subsec.~\ref{ssec:array_pipelines}.

The \ac{mvdr} beamformer exhibits low spatial selectivity for small $d$, leaving sources between $\pm\pi/3$ spatially unresolved while remaining free from aliasing effects. For large $d$, \ac{mvdr} shows improved selectivity, however, prominent sidelobes appear above $f_\mathrm{a}$ when the inter-microphone distance exceeds the Nyquist criterion, resulting multiple angles to have similar energy levels. The \ac{pf} provides modest improvements: for $d = \SI{1}{\centi\meter}$, it reduces the selectivity around \SI{0}{\radian} observed with \ac{mvdr}, while for $d = \SI{17}{\centi\meter}$, the sidelobes become less pronounced.

In contrast, JNF-SSF demonstrates consistent spatial selectivity patterns across both small and large $d$ values. For small $d$, despite the expected low spatial selectivity for closely spaced microphones, it achieves substantially better selectivity than the \ac{mvdr}-based frameworks. For large $d$, while sidelobes of similar shapes to those of \ac{mvdr} and \ac{mvdr}+\ac{pf} are present, they exhibit considerably lower energy compared to the look-direction. Our intuition regarding the superior spatial selectivity of joint nonlinear processing compared to separate processing stages lies in its ability to exploit interdependencies between spatial and tempo-spectral information.
Given the frequency-dependent nature of spatial aliasing and the fact that most spatial aliasing mitigation approaches analyze cross-frequency patterns, we believe that cross-frequency processing in the first \ac{lstm} layer of JNF-SSF is important for effective spatial aliasing mitigation.

\section{Conclusions}
\label{sec:conclusion}
In this work, we compared multichannel frameworks based on traditional linear \ac{mvdr} beamforming and \ac{dnn}-based joint nonlinear filtering in terms of their susceptibility to spatial aliasing and ability to maintain performance when inter-microphone distance exceeds the Nyquist criterion. Our analysis reveals fundamental differences in spatial aliasing robustness between these approaches. Both \ac{mvdr} and its concatenation with a \ac{pf}, where spatial and tempo-spectral processing are performed separately, suffer from prominent sidelobes in spatial selectivity patterns for large inter-microphone distance. In contrast, JNF-SSF that performs spatial and tempo-spectral processing jointly, demonstrates strong robustness to spatial aliasing conditions. It maintains both stable speech enhancement and consistent spatial selectivity patterns across different inter-microphone distances. We can conclude that joint processing of spatial and tempo-spectral information provides significant advantages over separate processing stages when dealing with spatial aliasing.

\balance
\bibliographystyle{IEEEbib_ref_initials}
\bibliography{refs25}

\end{document}